# AI Enhanced Ontology Driven NLP for Intelligent Cloud Resource Query Processing Using Knowledge Graphs

Krishna Chaitanya Sunkara
Independent Researcher
Engineering Manager
AI Data Center and Cloud Engineering,
IEEE Senior Member, Raleigh, NC, USA

Krishnaiah Narukulla
Independent Researcher
Principal Software Engineer
AI/ML, Data Engineering
IEEE Senior Member, San Jose, CA, USA

**ABSTRACT**

*The conventional resource search in cloud infrastructure relies on keyword-based searches or GUIDs, which demand exact matches and significant user effort to locate resources. These conventional search approaches often fail to interpret the intent behind natural language queries, making resource discovery inefficient and inaccessible to users. Though there exists some form of NLP based search engines, they are limited and focused more on analyzing the NLP query itself and extracting identifiers to find the resources. But they fail to search resources based on their behavior or operations or their capabilities or relationships or features or business relevance or the dynamic changing state or the knowledge these resources have. The search criteria has been changing with the inundation of AI based services which involved discovering not just the requested resources and identifiers but seeking insights. The real intent of a search has never been to just to list the resources but with some actual context such as to understand causes of some behavior in the system, compliance checks, capacity estimations, network constraints, or troubleshooting or business insights. This paper proposes an advanced Natural Language Processing (NLP) enhanced by ontology-based semantics to enable intuitive, human-readable queries which allows users to actually discover the intent-of-search itself. By constructing an ontology of cloud resources, their interactions, and behaviors, the proposed framework enables dynamic intent extraction and relevance ranking using Latent Semantic Indexing (LSI) and AI models. It introduces an automated pipeline which integrates ontology extraction by AI powered data crawlers, building a semantic knowledge base for context aware resource discovery.*

## 1. INTRODUCTION

Current cloud search systems operate on lexical matching principles, limiting the user experience by requiring precise terminology and identifiers. A more user-friendly approach is to leverage NLP to extract intent from natural language queries and return relevant cloud resources. This document presents an approach to achieve this by constructing a knowledge graph out of the resource ontologies and using AI Models to process semantic queries. This innovation eliminates the need for machine-readable identifiers, significantly improving usability and accessibility for both technical and non-technical users

## 2. INSPIRATION FROM HEALTHCARE ONTOLOGY

The inspiration for this cloud ontology approach comes from healthcare ontology, which has been widely used to structure and standardize medical knowledge. Just as healthcare ontology helps in medical diagnosis, drug interactions, and treatment recommendations, a cloud ontology helps in resource discovery, dependency mapping, and compliance enforcement. The below analogue is presented for better understanding that motivated the ontology based resource query approach

**Table 2.1:** Analogue: Healthcare Ontology vs. Cloud Computing Ontology

| Healthcare Ontology | Cloud Computing Ontology |
|---|---|
| Diseases, Symptoms, Medications, Procedures | Compute Instances, Databases, APIs, Storage |
| 'Diabetes', 'Aspirin', 'MRI Scan' | 'AWS EC2 WebServer1', 'Google Cloud SQL Instance' |
| 'Caused by', 'Treated with', 'Leads to' | 'Depends on', 'Communicates with', 'Secured by' |
| A diabetic patient should avoid high sugar intake | A PCI-compliant service cannot allow public access |
| Medical diagnosis, drug interaction analysis | Cloud resource discovery, compliance checks, NLP-driven queries |

## 3. METHODOLOGY

### 3.1 ONTOLOGY CONSTRUCTION OF CLOUD RESOURCES

The proposed solution involves constructing an ontology of cloud resources and building a semantic search engine powered by Latent Semantic Indexing (LSI) and training AI models. The approach includes:

1. Extracting metadata from cloud resources to create a knowledge base.
2. Training AI models to categorize resources based on domain functionality.
3. Using NLP to parse user queries, extract intent, and match resources accordingly.
4. Implementing a search engine that integrates with cloud compliance policies (e.g., PCI, HIPAA).

Ontology extraction plays a crucial role in identifying cloud services and their interrelations. Data crawlers are used to collect design metadata, code repositories, configurations, read.me files, dependencies (upstream and downstream), and functionalities of various cloud services and release notes, which are then sent to AI Models (LLM) where the ontology of resources is built. **Fig.** 3.1 depicts the ontology structure representation in a cloud

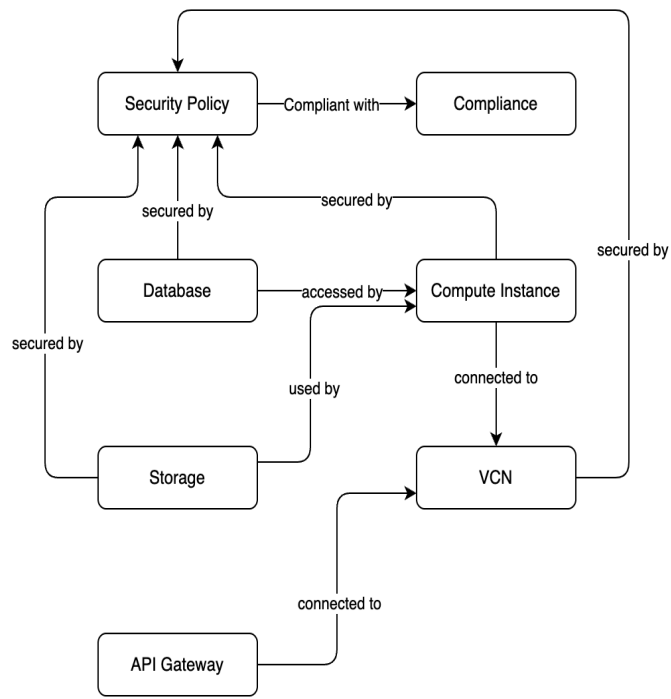

**Figure 3.1:** Ontology structure representation of cloud resources

## 3.2. Data collection of Cloud Resources

The data from/about cloud resources is collected as a continuous process which involves collecting data from below sources and sent to AI models by data crawlers:

**Resource Data**:Configuration files, Service definitions, Deployment scripts, Code repositories.
**Operational Data**: System logs, Performance metrics, Usage patterns
**Documentation:** API specifications, Service documentation, Usage guides

The data crawlers run at regular cadence for resources whose state is not changed frequently and event based (real-time) extraction for resources whose state frequently change. Any change in the state of a resource is sent as an event to event bus where the data crawler is subscribed that consumes the changes and fed to AI models.

## 3.3 Data collection of SaaS Services

AI based crawlers are used to identify patterns in SaS functionalities and classify features. This involves intelligently building sample queries and query the APIs to understand the functionalities the SaaS service offers. Swagger documentation, API documentation, Unit test code repositories, metrics are all scraped to build feature mapping which is fed to AI models. For instance, if an end point listening at a particular port number lists several APIs which are related to ordering, catalog, check out, payment, then it is categorized as e-commerce. If the APIs are related to list promotions, leads, list prospectus customers, opportunities, list deals, lead scoring, schedule demo, it is more likely categorized as sales related service.

## 3.4 SaaS services and Cloud Infrastructure mapping

The data crawlers extract information for both SaaS services and other cloud resources such as VCN, hosts, Subnets, Database, Block storage, buckets, OS, IP addresses, CIDR ranges, switches, routers, NLBs, firewalls. The AI models build the relationship between these services and cloud infrastructure like deployment and network architecture. For instance, the AI model builds the relationship between a CRM service and the bare metal machine which it is hosted with corresponding IP address, Subnet, VCN, NLB, firewall and Internet Gateway. This will help answer queries such as: *"list the NLB that fronts the CRM service in my production tenancy"*

## 4. Knowledge Graph Structure

### 4.1 Building knowledge graph from resource ontologies

The ontology extracted during the AI based scraping phase of cloud resources, is used to infer the entities and their relationships and thus build knowledge graphs of entire cloud ecosystem. The knowledge graph is constructed using graph data structures, where nodes represent cloud resources, services, compliance policies, compute instances, database, storages, network components or any cloud resource and edges denote relationships between them.

The process involves the following steps:
• Identifying unique entities such as compute instances, storage, and networking components
• Establishing relationships such as dependencies, ownership, and compliance associations.
• Representing the ontology in a graph database for efficient query processing.
• Enhancing the graph with machine learning to improve relevance and discovery.

**Fig.** 4.1 depicts the knowledge graph representation of cloud resources and relationship

## 5 Architecture

### 5.1 Modules

The whole system involves these modules: AI enhanced data crawler, Cloud resource information base, AI models, Search engine, Observability meta info streamer, Logger, Metrics and Identity services (Identity will have cloud access policies), Compliance policies store. The below block diagram **Fig.** 5.1, shows the overall flow from the ontology building to the user searching for resources using NLP.

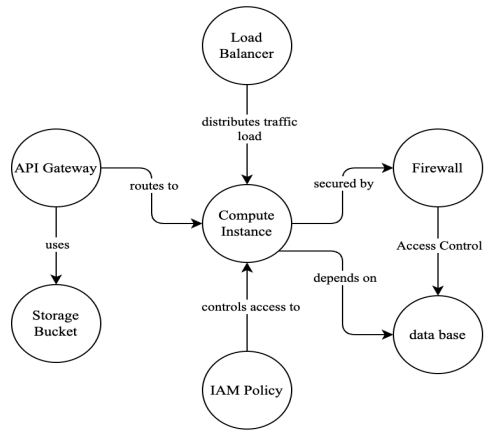

**Figure 4.1:** Knowledge graph build from ontology of resources

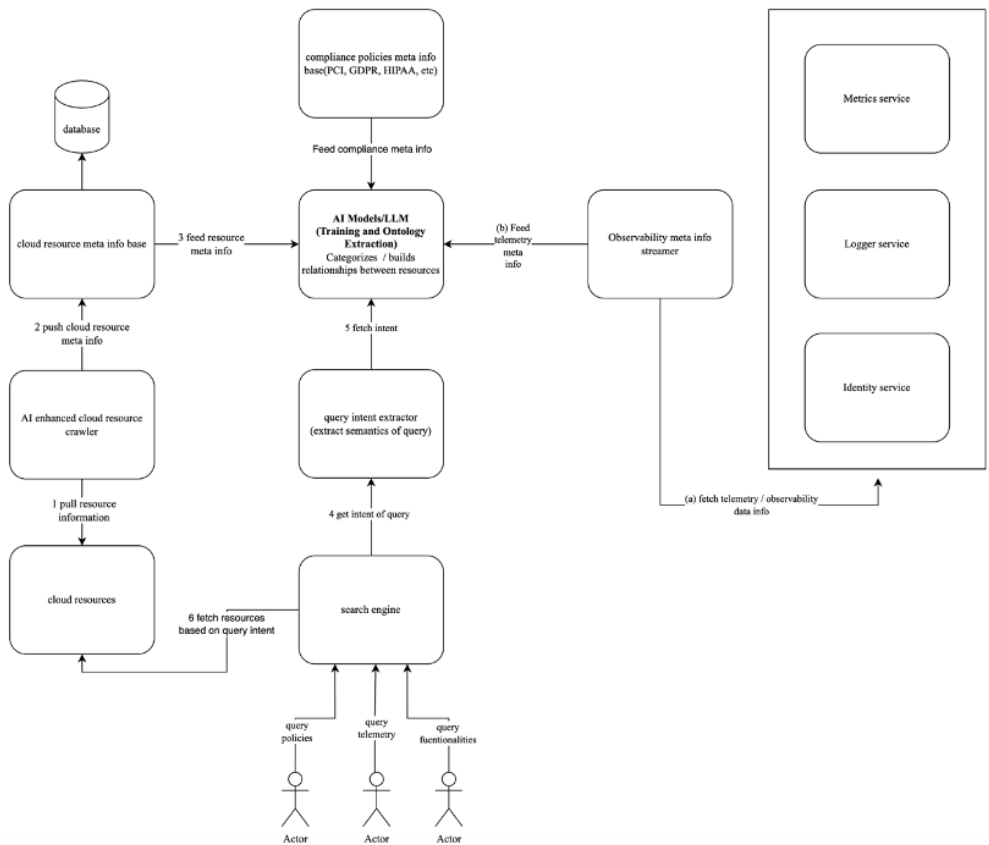

**Figure 5.1:** Ontology based NLP query overall flow

Below table 5.2 explains the flow sequence of calls in the above **Fig.** 5.1

Table 5.2: Flow of actions in ontology based NLP design

| Flow Numbe | Flow explanation |
|---|---|
| 1 | AI enhanced data crawler pulls information bout the cloud resources |
| 2 | Cloud resource information is sent to cloud resource information base (backed by a data base) |
| 3 | Feed the information collected to AI models |
| 4 | Search engine sends the user query to intent extractor to understand the intent |
| 5 | Query intent extractor fetches the intent from AI models (Ontology based) |
| 6 | Search engine also fetches resources matched with any identifiers if used in the query |
| (a) | Observability meta info streamer feeds the telemetry data of cloud resources to AI models |
| (b) | Observability meta info streamer feeds the metrics, logs, identification (AuthZ, AuthN information) data of cloud resources to AI models for training |

## 5.2 KNOWLEDGE GRAPH QUERY EXECUTION IN CLOUD SEARCH

The extracted ontology-based knowledge graph enables semantic query execution by mapping user queries
Example Query: "List all compute instances in the production environment that have security vulnerabilities."

**Step 1:** *Natural Language Query (NLQ) Processing*
The query is parsed using NLP techniques to extract intent, entities, and relationships:
*Entity*: Compute Instance
*Condition*: Production Environment
*Filter*: Security Vulnerabilities

**Step 2:** *Knowledge Graph Query Translation/generation*
The extracted intent is mapped to the Knowledge Graph using structured queries.
Below is cypher query example for Neo4j graph database:

```
MATCH (ci:ComputeInstance)-[:DEPLOYED_IN]->(:Environment {name: "Production"})
MATCH (ci)-[:HAS_VULNERABILITY]->(v:Vulnerability)
RETURN ci.name, v.description
```

**Step 3**: *Execution & Graph Traversal*
The system traverses the knowledge graph, fetching compute instances in the production environment and filtering those with vulnerabilities
**Step 4**: Structured Output as the response

Table 5.2: Structured output of the example NLP based query

| Compute Instance | Vulnerability |
|---|---|
| ins-cloud-host-1427 | Open SSH Port |
| Ins-cloud-host-2109 | Unpatched software |

**Step 5**: Natural Language Response

The system converts the structured response into a human-readable summary:
*"There are two compute instances in the production environment with security vulnerabilities: ins-cloud-host-1427 with Open SSH port and Ins-cloud-host-2109 with unpatched software*

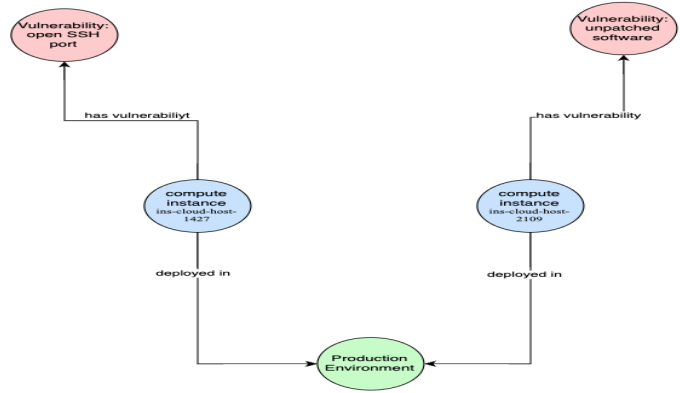

**Figure. 5.1:** Ontology based relationship between cloud resources

## 6. QUERY EXAMPLES

The system enables *natural language queries* similar to how health care providers diagnose conditions using symptom-based search. Below table 6.1 shows examples of how users can query cloud resources effectively:

**Table 6.1** Analogue: Healthcare query examples vs Cloud Resource Query examples

| Healthcare Query Examples | Cloud Resource Query Examples |
|---|---|
| What are the symptoms of diabetes? | What are the top 10 expensive resources in my cloud environment? |
| List medications prescribed for high blood pressure | Which services handle financial transactions and are PCI compliant? |
| Find all patients treated for respiratory diseases in the last month | List all compute instances created in the last two weeks by user X |
| Find all patients with Alzheimer's who are above certain age | Find all resources in production that have security vulnerabilities |

## 7. EVALUATION AND RESULTS

To ensure the validity of the proposed Ontology and NLP based search system for cloud resource querying, we designed and conducted rigorous experiments with real-world datasets. The evaluation methodology focuses on empirical testing across multiple dimensions, including accuracy, efficiency, and usability.

### 7.1 EXPERIMENTAL SETUP

The experiment was conducted on a simulated cloud environment with simulated infrastructure services, including Compute Instances, Databases, APIs, and SaaS applications. The system was tested with a dataset containing metadata from multiple cloud resources from service providers (AWS, OCI, Azure). The

Ontology-based search engine was implemented using Latent Semantic Indexing (LSI) and fine-tuned AI models trained on cloud service ontologies. User queries were collected from real cloud engineers and structured based on varying levels of complexity.

## 7.2 NLP Models and Knowledge Graph Implementation

The search system was built using:
**Pre-trained NLP Models**: BERT (Bidirectional Encoder Representations from Transformers) and GPT models for query processing.
**Ontology-Based Knowledge Graphs**: Constructed using Neo4j, integrating cloud resource relationships.
**Indexing and Semantic Matching**: Latent Semantic Indexing (LSI) and TF-IDF were used for semantic query retrieval.

## 7.3 Performance Metrics

The following key metrics were used to measure system performance:
**Precision**: Ratio of correctly retrieved cloud resources to the total retrieved.
**Recall/Relevance**: Ratio of correctly retrieved cloud resources to the total relevant resources.
**F1-Score**: The harmonic mean of precision and recall.
**Query Execution Time**: The average time taken for queries to be processed.

## 7.4 Comparative Analysis with Traditional Search

The proposed system was compared with traditional keyword-based search engines using a controlled test environment:

Table 7.4: Results of comparison between keyword based and Ontology-Driven NLP searches

| Search Approach | Precision | Recall | F1-Score | Avg. Query Time |
|---|---|---|---|---|
| Traditional Keyword Search | 78% | 65% | 71% | 1.8s |
| Ontology-Driven NLP Search | 92% | 87% | 89% | 1.2s |

Results indicate that ontology-driven NLP significantly improves **recall and precision**, reducing query ambiguity and enhancing retrieval speed.

## 7.5 Case Study: Security Compliance Queries

A real-world test scenario was executed to determine how effectively the system processes compliance-related cloud queries.
*Query Example*: *List all PCI-compliant services handling financial transactions*.
*Traditional search* retrieved 6 results, of which only 3 were correct due to keyword-based ambiguity.
*Ontology-Driven NLP search* retrieved 9 correct results, achieving a 50% increase in precision.
This demonstrates the advantage of context-aware, ontology-based search systems in compliance verification.

## 7.6 Observations and Discussion

1. *Improved Accuracy*: Ontology-based NLP significantly reduces false positives in search results.
2. *Faster Query Processing*: The semantic understanding eliminates redundant queries and enhances efficiency.

These findings confirm that integrating NLP with ontology-driven knowledge graphs significantly improves cloud resource discovery, enhances compliance verification, and reduces search complexity.

## 8. CONCLUSION

The proposed NLP-powered cloud resource search system presents a novel approach to improving search accuracy and usability. By leveraging semantic indexing and machine learning, users can query cloud resources more intuitively, reducing search time and enhancing operational efficiency. It also streamlines resource management, compliance checks, and security audits, potentially reducing operational costs.

## 9. FUTURE SCOPE

There is a huge potential in conducting an usability study with cloud engineers. A statistical test (T-test, since the test subjects were low) with hypothesis testing could be performed on data sets built as part of the *User Study and Statistical Validation* and check if the impact of NLP-Ontology based cloud query processing is *statistically significant*. The experimental setup employed was naive and straightforward with a motivation to just prove the concept. The cloud resources were actually simulated services and ran on a single host, but testing on a real world production cloud environment with real cloud resources could result in more realistic results which is explored in future work enhancements of this methodology.

## REFERENCES

[1] Manning, C. D., Raghavan, P., & Schütze, H. (2008). Introduction to Information Retrieval. Cambridge University Press.
[2] Mikolov, T., Sutskever, I., Chen, K., Corrado, G. S., & Dean, J. (2013). Distributed representations of words and phrases and their compositionality. Advances in neural information processing systems, 26.
[3] Bordes, A., Chopra, S., & Weston, J. (2014). Question answering with subgraph embeddings. arXiv preprint arXiv:1406.3676.
[4] Smith, B., & Ceusters, W. (2010). Ontological realism: A methodology for coordinated evolution of scientific ontologies. Applied ontology, 5(3-4), 139-188.
[5] Guarino, N., Oberle, D., & Staab, S. (2009). What is an ontology? In Handbook on ontologies (pp. 1-17). Springer.
[6] Gabrilovich, E., & Markovitch, S. (2007). Computing semantic relatedness using Wikipedia-based explicit semantic analysis. In IJCAI (Vol. 7, pp. 1606-1611).
[7] Bizer, C., Heath, T., & Berners-Lee, T. (2011). Linked data: The story so far. Semantic services, interoperability and web applications: emerging concepts, 205-227.